\documentclass[intlimits,twoside,a4paper]{article}

\usepackage{amsmath,amssymb}
\usepackage{graphicx}
\usepackage{float}
\usepackage[T2A]{fontenc}
\usepackage[cp1251]{inputenc}

\usepackage{wrapfig}

\usepackage{color}

\usepackage{cmpj2}

\issue{2016}{19}{2}{23802}
\doinumber{10.5488/CMP.19.23802}

\title[The contribution of electrostatic interactions to the collapse of oligoglycine in water]{The contribution of electrostatic interactions to the collapse of oligoglycine in water}

\author[D. Karandur, B.M. Pettitt]{D. Karandur\refaddr{label1}\thanks{Present address: Howard Hughes Medical Institute at the University of California, Berkeley, CA 94720, USA}\,,
B.M.  Pettitt\refaddr{label1,label2}\thanks{E-mail: mpettitt@utmb.edu}}
\addresses{
\addr{label1}Structural and Computational Biology and Molecular Biophysics, Baylor College of Medicine, \\ Houston, TX 77030, USA
\addr{label2}Sealy Center for Structural Biology and Molecular Biophysics, Department of Biochemistry and Molecular Biology, University of Texas Medical Branch, 301 University Blvd., Galveston, TX 77555-0304, USA
}

\date{Received November 3, 2015, in final form January 11, 2016}
\authorcopyright{D. Karandur, B.M. Pettitt, 2016}

\begin{document}

\maketitle


\begin{abstract}
Protein solubility and conformational stability are a result of a balance of interactions both within a protein and between protein and solvent. The electrostatic solvation free energy of oligoglycines, models for the peptide backbone, becomes more favorable with an increasing length, yet longer peptides collapse due to the formation of favorable intrapeptide interactions between CO dipoles, in some cases without hydrogen bonds. The strongly repulsive solvent cavity formation is balanced by van der Waals attractions and electrostatic contributions. In order to investigate the competition between solvent exclusion and charge interactions we simulate the collapse of a long oligoglycine comprised of 15 residues while scaling the charges on the peptide from zero to fully charged. We examine the effect this has on the conformational properties of the peptide. We also describe the approximate thermodynamic changes that occur during the scaling both in terms of intrapeptide potentials and peptide-water potentials, and estimate the electrostatic solvation free energy of the system.
\pacs 87.15.ap, 87.10.E-, 87.15.Cc
\end{abstract}


\section{Introduction}

The polypeptide chain of proteins collapse in aqueous solvent due to a complex interplay of correlations or effective interactions within the protein, and between the protein and solvent. Salt effects can stabilize the fold, or destabilize it, based on the overall electrostatics and the interplay with solvent forces~\cite{Smith1991}. Similar mechanisms have been observed to cause changes in peptide and protein solubility as well \cite{Cohn1943, Arakawa1987,Perkyns1996}. Electrostatically driven reordering  is not restricted to the solvent about a protein or peptide solute but has been observed in simulated aggregates of pentaglycines in water \cite{Karandur2014a}. In intrinsically disordered proteins, different  arrangements of charged amino acids have been observed to lead to changes in solubility and conformation \cite{Mao2010, Muller-Spath2010}.

The transfer model can be used to qualitatively describe the changes in free energy when the solubility or structure of a protein changes with respect to a solvent \cite{Tanford1964}. Bolen and co-workers used a variant of the transfer model to show that backbone-solvent interactions are a major contributor to the free energy changes during protein folding \cite{Auton2007, Auton2011}. There remain issues which make a precise quantitative argument for the contribution of the backbone versus side chains based on the transfer model difficult \cite{Dill1997,Auton2011,Canchi2011,Moeser2014}.

Oligoglycine is a  tested model of the protein backbone \cite{Tran2008, Hu2010a, Hu2010b, Teufel2011}. Previous work has shown that the electrostatic solvation free energy for oligoglycines at infinite dilution becomes more favorable with an increasing chain length \cite{Hu2010b} while an isolated peptide collapses as seen both experimentally \cite{Teufel2011} and in simulation \cite{Tran2008, Hu2010a} to interact with itself. We have previously shown that this collapse is due to a balance between non-hydrogen bonding dipolar correlations (so-called ``CO-CO'' interactions) and solvent cavity formation \cite{Karandur2016}. If we decompose the solvation free energy for peptides into the underlying force field components of van der Waals (vdW) cavity formation and electrostatics recent results show that the vdW free energy does not follow naive hydrophobicity arguments \cite{Kokubo2011, Kokubo2013, Harris2014}. The cavity has a repulsive component which is unfavorable but in many cases the vdW attractions more than compensate and produce an uncharged cavity which has a favorable free energy when the surface area is maximized as in a strictly linear peptide without electrostatic contributions \cite{Kokubo2013, Harris2014}. However, when peptides of alanine were allowed to conformationally relax the vdW attractions no longer outweighed the repulsive cavity formation \cite{Kokubo2013}.

The vdW terms for such solutes with strong dipoles like glycine in a polar solvent like water are relatively small. Using the common classical force field components, electrostatics contribute the most to the solvation free energy \cite{Hu2010b}. Several methods to estimate the solvation free energy are in the literature \cite{McCammon1987, Brooks1988}. Approximate methods based on mean field approaches such as Poisson Boltzmann theory \cite{Gilson1987, Sharp1987} and Generalized Born theory \cite{Still1990} are relatively inexpensive, but are based on the assumption that the solvent is a dielectric continuum and responds linearly to solute electrostatics. While near linearity has been observed for systems comprised of monovalent ions \cite{Born1919, Jayaram1989}, complex systems such as proteins have complicated multipolar distributions and show marked deviations from linearity \cite{BinLin2010}. Furthermore, at length scales equivalent to the size of water molecules, there is evidence that continuum models break down \cite{Blaak2006}. Hybrid models, where the solute and a solvation shell are modelled explicitly while the rest of the solvent is modelled by a continuum representation, overcome some of these issues \cite{Okur2006, Lin2009, Lin2011}. More accurate are methods such as thermodynamic integration and free energy perturbation calculations \cite{Tembre1984, Straatsma1991, Chipot1994} using molecular dynamics with explicit descriptions of solute and solvent atoms. While these methods are much more expensive, requiring extensive sampling, they are, in principle, capable of being much more accurate \cite{BinLin2010, BinLin2011}.

Recent work considered the change in the vdW cavity free energy between extended and collapsed conformations for oligoalanine \cite{Kokubo2013} in relation to the changes in solvent exposed area and classical hydrophobicity \cite{Pratt1986}. Given the cavity contribution to oligoglycine \cite{Hu2010b}, alanine \cite{Kokubo2013} and other peptide/pro\-tein collapse in water, we wish to characterize the electrostatic contributions and show how they change the conformational manifold. In order to consider the contribution of the electrostatic solvation free energy to a flexible peptide we have chosen Gly$_{15}$, to compare with the previous, related work \cite{Hu2010b}. We perform a series of simulations scaling charges on the peptide from zero to fully charged. We examine the effect that the scaling charges have on peptide conformation. We then examine other structural and thermodynamic changes both within the peptide and between the peptide and water as charges are turned on. The next section describes the models and methods used. We then describe the results in terms of the effects that the variations in potential have on the structure followed by a detailed examination of the accompanying thermodynamic changes, and end with our conclusions.

\section{Methods}

An oligoglycine peptide  15 residues (Gly$_{15}$) with capped ends was built with CHARMM \cite{CHARMM2009}, in the fully extended state. The peptide was solvated with TIP3P water with VMD \cite{Humphrey1996} such that there are at least 5 solvation shells about the extended peptide.

We will consider the amount of the charge potential added to the peptide solute as a scaling factor $\lambda$. The electrostatic contribution to the solute-solvent potential, $U^\text{elec}(r)$  is thus linearly scaled as $\lambda U^\text{elec}(r)$. For a sufficient number of $\lambda$ points and sampling at each point, this would constitute a free energy charging coordinate.

At constant pressure and temperature, the change in free energy of solvation with respect to adding charges to the solute peptide,
\begin{equation}
\label{eq:1}
\Delta G = \int_0^1 \langle U^\text{elec} \rangle_{\lambda} \rd\lambda,
\end{equation}
where $\langle \ldots \rangle_{\lambda}$ denotes the configurational average at a particular value of $\lambda$.

Here, we used six $\lambda$ windows  to scale the electrostatics: 0.0, 0.2, 0.4, 0.6, 0.8, 1.0. While the electrostatics in some systems yields to linear response theory, a more complicated integrand with curvature means that more quadrature points may be required depending on the precision required. In addition, sampling at each point for flexible systems is notoriously slow \cite{Kokubo2013}. For these reasons, the use of equation (\ref{eq:1}) here results in only a rough approximation. We wish to consider the competition between the cavity formation and the charging on the conformational distribution. The range of $\lambda$ points chosen will allow us to consider the structural measures of the system between a purely uncharged but flexible vdW cavity and the fully charged force field.

All the systems were minimized for 25~K steps with the peptide held fixed and then for 25~K steps where all the molecules are allowed to move. They were then equilibrated for 2~ns in the $NPT$ ensemble (pressure $= 1$~atm, temperature $= 300$~K), followed by 1~ns in the $NVT$ ensemble. Systems with $\lambda = 0.0$, 0.2, 0.4 and 0.6 were then run for 300~ns in the $NPT$ ensemble. Systems with $\lambda = 0.8$ and 1.0 were run for 500~ns. The longer runs at these $\lambda$ points allowed a more extensive sampling because the charged interactions, both within the peptide and between the peptide and water, affect the conformational manifold to a greater extent. The NAMD package was used to run the simulations \cite{Phillips2005} with the CHARMM27 force field \cite{CHARMMFF1998, CHARMMFF2004} with a timestep of 1~fs. Particle Mesh Ewald with a grid size of 1~{\AA} was used to calculate the long-range electrostatics \cite{Darden1993}. The RATTLE module was used to constrain all hydrogen bonds \cite{Andersen1983}.

Components of the approximate free energy for our process corresponding to the response of other terms in the potential energy model require doing integrals similar to the one presented above. We decomposed the potential energy both within the peptide and between the peptide and water into standard bonded (only intrapeptide) and non-bonded terms. The NAMD Energy plug-in was used to calculate these terms \cite{Phillips2005}. Particle Mesh Ewald was not used during the calculation of these various potentials.

\section{Results}

\subsection{Gly$_{15}$ collapses in a charge-dependent manner}

\begin{figure}[!b]
\centering
\includegraphics[width=0.55\textwidth]{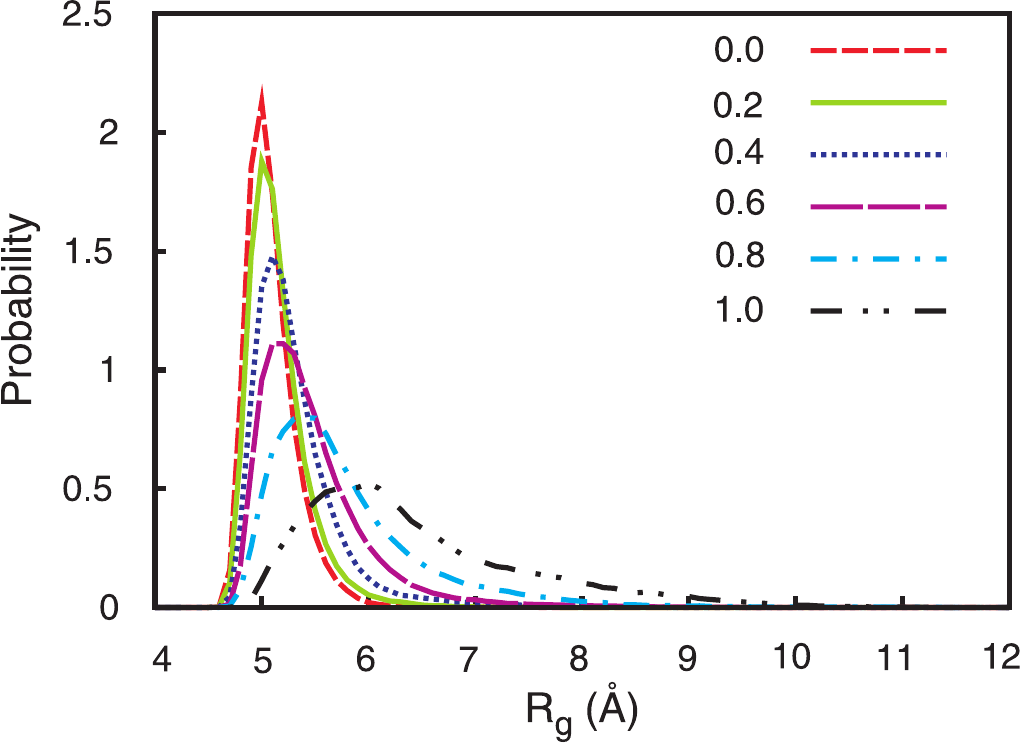}
\caption[Probability distribution of radius of gyration with respect to $\lambda$.]{(Color online) Probability distribution of radius of gyration of peptide for different $\lambda$ windows. The color scheme for each $\lambda$ is described in the key.}
\label{figure1}
\end{figure}

The peptide conformational distribution  collapsed in a few nanoseconds at each $\lambda$. The systems explored different distributions with respect to $\lambda$, as will be described in detail below. In all the systems, they explored a collapsed set of conformers versus a random chain. The systems rarely explored extended conformations, usually for a few ns or less, before collapsing again. Figure~\ref{figure1} shows the distribution of the radius of gyration in each of the different charge state windows. The system with all charges turned off, i.e., where only vdW interactions both within the peptide and between the peptide and water are possible, shows the narrowest distribution. Since  peptides normally strongly interact with water via dipolar hydrogen bonds, turning the charges off prevents these interactions, allowing the peptide to collapse via intrapeptide van der Waals interactions and via the cavity formation forces even though the intermolecular vdW attractions oppose this \cite{Kokubo2013}.

\begin{figure}[!t]
\centering
\includegraphics[width=0.8\textwidth]{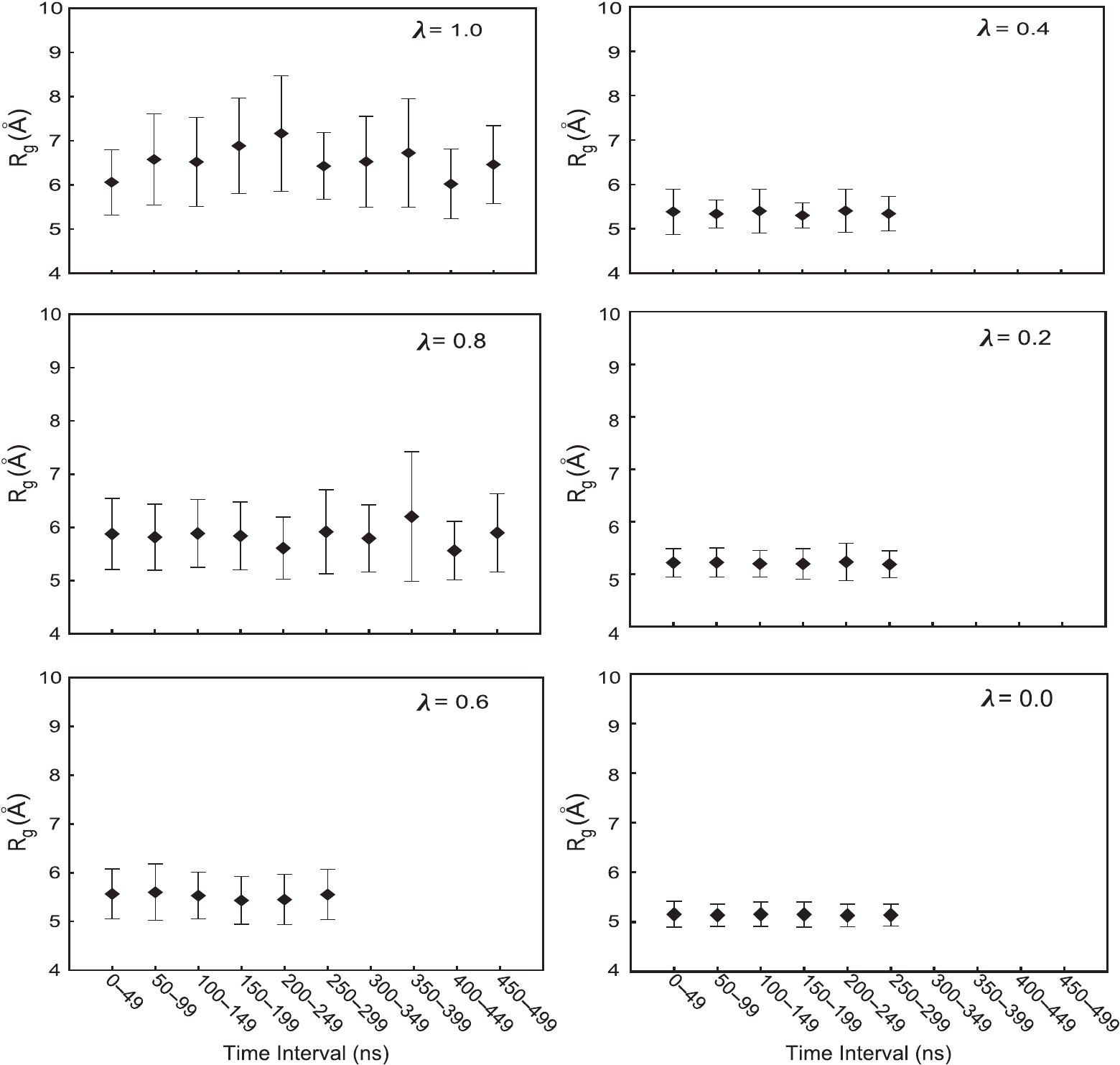}   
\caption[Radius of gyration across time with respect to $\lambda$.]{Variation in the radius of gyration of peptide across time for different $\lambda$ windows.}
\label{figureS1}
\end{figure}

\begin{figure}[!t]
\centering
\includegraphics[width=0.7\textwidth,height=0.51\textheight]{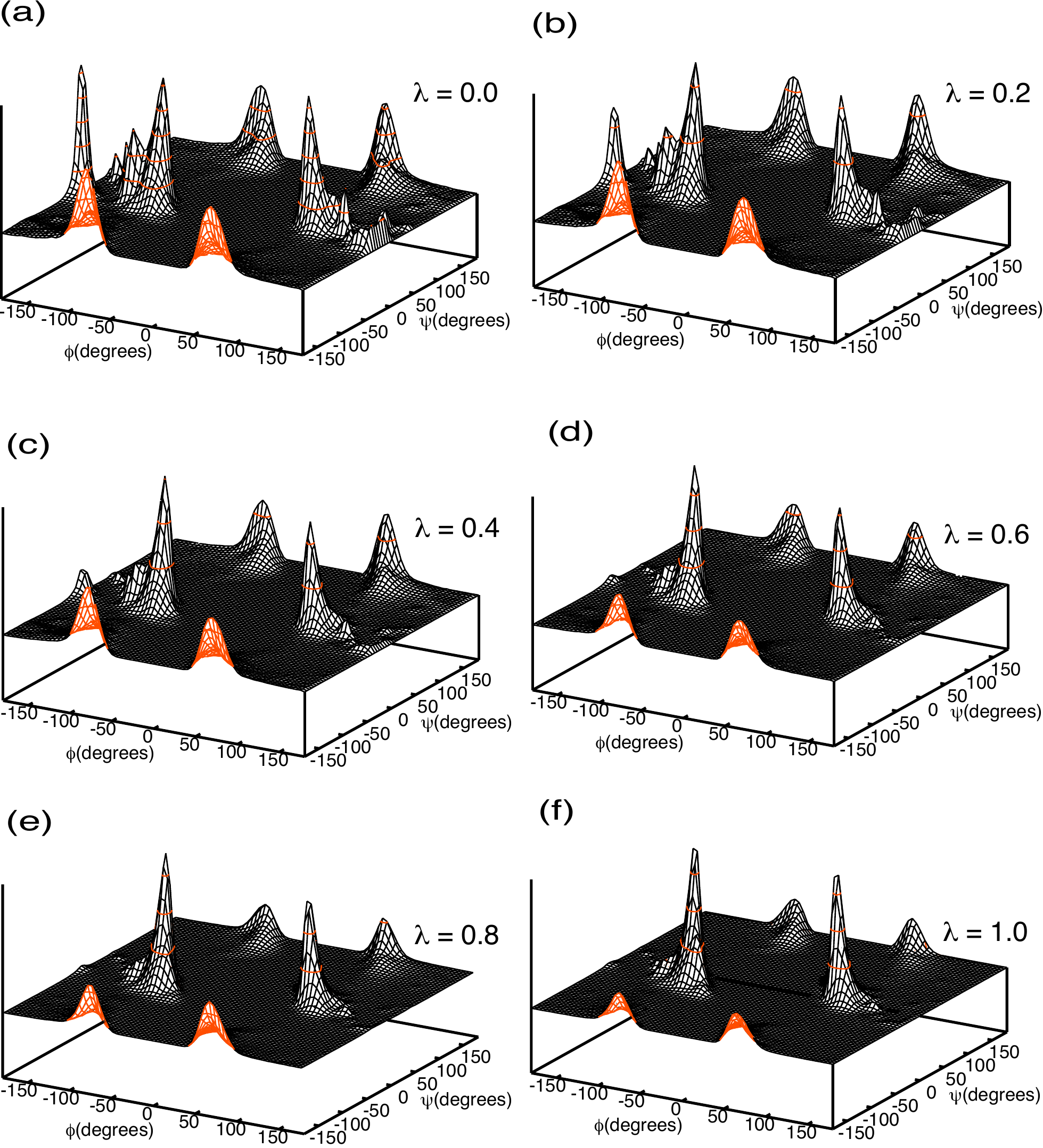}   
\caption[Ramachandran Plots with respect to $\lambda$.]{(Color online) Ramachandran plots of Gly$_{15}$ with respect to $\lambda$.}
\label{figureS2}
\end{figure}

As charges are turned on, the distribution broadens, since the charge-charge peptide-water interactions allow the peptide to explore a larger range of conformations. This is also reflected in the increase in the mean radius of gyration with respect to an increasing charge. The radius of gyration also takes longer to converge  (figure~\ref{figureS1}). Thus, the higher $\lambda$ value systems show much more variation whereas the smaller $\lambda$ value systems apparently converged within a 100~ns. Furthermore, the peptide in all the systems explored  the structural conformations of  oligoglycine with different probabilities (figure~\ref{figureS2}). At lower $\lambda$ values, the peptide explores all the allowed regions of the Ramachandran space with almost similar probabilities, but as the charges are turned on, the peptide becomes more constrained to the regions that are consistent with that of glycines for this particular force field \cite{CHARMMFF2004}.

\subsection{Potential average components with respect to charge}

\begin{figure}[!h]

\centering
\includegraphics[width=0.95\textwidth]{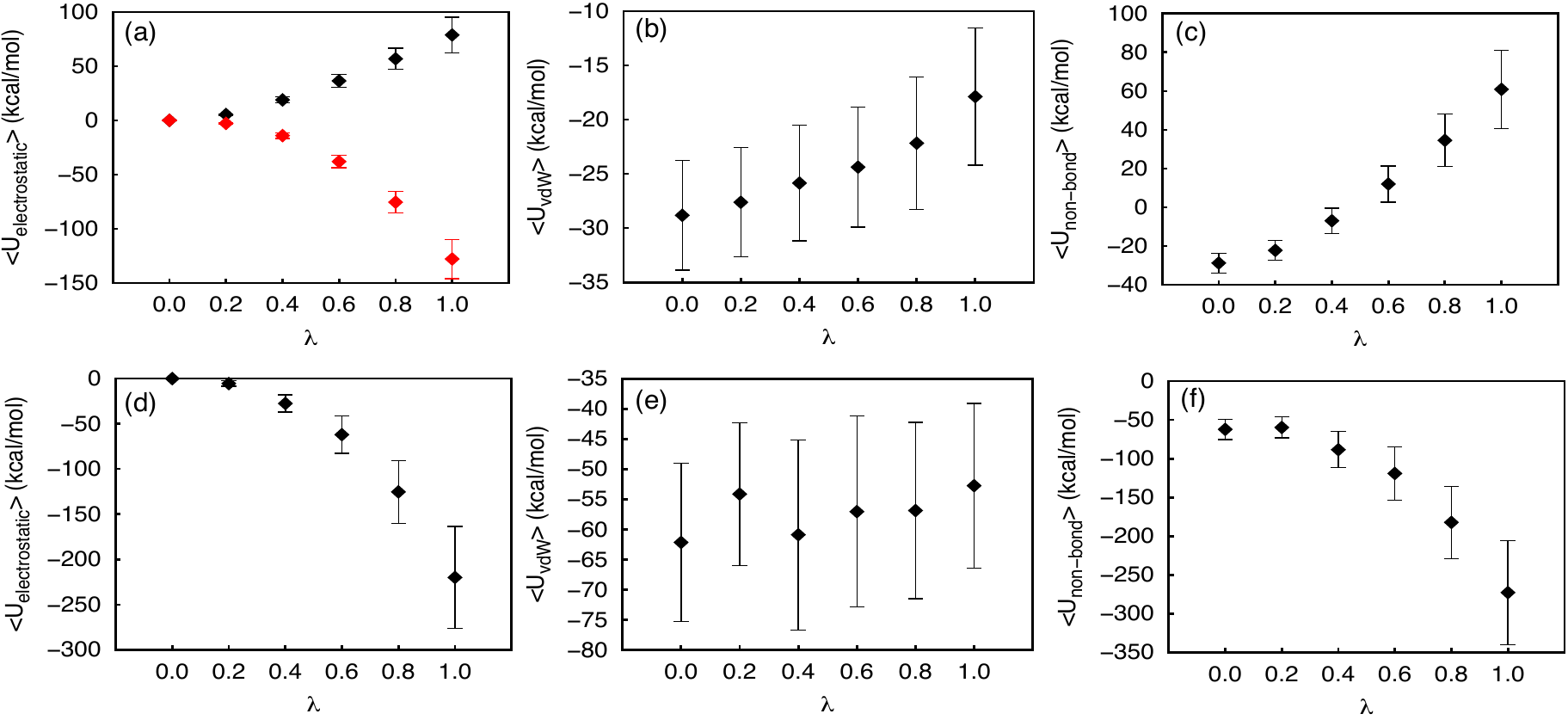}
\caption[Potential energy with respect to $\lambda$ (non-bonded).]{(Color online) Average potential energy with respect to $\lambda$. (a) shows intrapeptide electrostatic potential energy. The total electrostatic potentials are shown in black and electrostatic potentials when excluding $\text{C}\cdots\text{C}$ and $\text{O}\cdots\text{O}$ interactions between atoms of adjacent CO dipoles are shown in red. (b) shows intrapeptide van der Waals potentials. (c) shows total intrapeptide non-bonded potential. (d) shows peptide-water electrostatic potentials, (e) shows peptide-water van der Waals potentials and (f) shows total peptide-water non-bonded potentials.}
\label{figure2}
\end{figure}

We decomposed the non-bonded potential averages as a function of $\lambda$ into electrostatic and van der Waals components for both intrapeptide and peptide-water potential energies. The results are shown in figure~\ref{figure2}. Both intrapeptide electrostatic and van der Waals potentials increase with an increasing $\lambda$, indicating that as the charges are turned on, intrapeptide interactions become unfavorable [figures~\ref{figure2}~(a) (points in black) and \ref{figure2}~(b)]. We have previously observed that oligoglycine collapse is not stabilized by intrapeptide H-bonds \cite{Hu2010a, Karandur2016}, but by a large number of dipole-dipole or ``CO-CO'' interactions. However,  the unfavorable nature of the intrapeptide electrostatic potential as the charge increases requires more analysis. We and others have previously observed that a strong positive correlation occurs between adjacent peptide dipoles at 3~{\AA}, which is inherently present in all polypeptides \cite{Bartlett2010,Karandur2014a, Karandur2016}. We note that gamma turns or C$_7$ rings, which would have favorable near-neighbor dipoles, are extremely rare in proteins or peptides in solution \cite{Pettitt1985}. In the more probable orientations, the near-neighbor dipoles are more aligned with each other and form locally strong but unfavorable interactions. We also note that the proximity of the near-neighbor dipoles is a requirement of forming a covalently linked polypeptide, and so becomes a source of frustration in the energy landscape \cite{Bryngelson1987}.

We now consider the rest of the electrostatics excluding the aspects of the CO-CO interaction surface associated with the near-neighbor amide dipoles. In order to exclude this near-neighbour effect, we calculated the intrapeptide potential while excluding $\text{C}\cdots\text{C}$ and $\text{O}\cdots\text{O}$ interactions between the atoms on the adjacent CO dipole atom pairs. The results are shown in figure~\ref{figure2}~(a) (points in red). These interactions alone account for much of the positive component of the intrapeptide electrostatic potential. The entire intramolecular electrostatic potential becomes favorable at higher $\lambda$ values without the near-neighbor repulsions.

The intrapeptide vdW potentials increase as the charges increase due to favorable electrostatic interactions inducing the changes in peptide conformational manifold with stronger contacts. It is interesting to note that while the electrostatic potentials show an increased variation about the mean with an increasing $\lambda$, the vdW potential averages show the same spread irrespective of $\lambda$. This leads to the total non-bonded contributions variances as well as means being dominated by the electrostatics.

As charges are turned on, the peptide-water electrostatic interactions become increasingly favorable [figure~\ref{figure2}~(d)], indicating the formation of favorable, dipolar, peptide-water interactions such as hydrogen bonds. As these interactions become available to the systems, they tend to add stability to the extended conformations that these systems explore. As noted above, these systems do not assume extended conformations with a high probability. As observed previously \cite{BinLin2010}, the peptide-water potential shows deviations from linearity in $\lambda$. The peptide-water vdW potential does not show as much variation with respect to $\lambda$ [figure~\ref{figure2}~(e)]. Thus, the total non-bonded intrapeptide potential increases as charges are turned on, whereas total non-bonded peptide-water potential energy becomes more favorable [figures~\ref{figure2}~(c) and \ref{figure2}~(f)], and in opposition they induce the peptide to explore larger conformational states.

\begin{figure}[!h]
\centering
\includegraphics[width=0.95\textwidth]{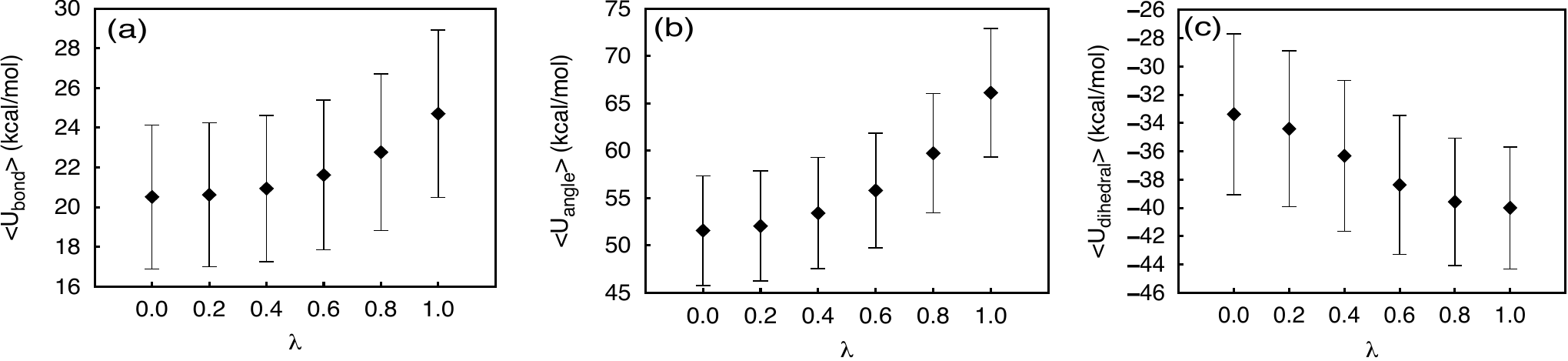}
\caption[Potential energy with respect to $\lambda$ (bonded).]{Potential energy with respect to $\lambda$ for (a) intrapeptide bond potential energy, (b) intrapeptide angle potential energy and (c) intrapeptide dihedral potential energy.}
\label{figureS3}
\end{figure}

We also considered whether the structural changes induced by turning the charges on affected the intrapeptide bonded potential energies (figure~\ref{figureS3}). Both the intrapeptide bond energy and the angle energy become slightly unfavorable from the induced strain as the charges are turned on. However, the dihedral angle energy becomes a bit more favorable with an increasing $\lambda$ [figure~\ref{figureS3}~(c)] indicating that the peptide is assuming conformations that will stabilise it in charged environments. Apparently, the more compact conformations that the peptides assume at lower $\lambda$ values, are unfavorable in terms of the dihedral potential.

\subsection{Approximate electrostatic contributions to free energy}

The approximate electrostatic solvation free energy was estimated as described in the methods as -66 kcal/mol. Previous calculations on short oligoglycines have estimated the electrostatic solvation free energy of pentaglycines to be $-42.53$~kcal/mol for the extended form \cite{Hu2010b} and $-37.3$~kcal/mol for the helical form \cite{Staritzbichler2005}. Since the solvation free energy per peptide is additive with respect to the number of residues, as was shown to be true for the extended and collapsed forms \cite{Kokubo2013}, the expected solvation free energy contribution for Gly$_{15}$ should be $\sim -127$~kcal/mol using the linear conformation estimate or $-112$~kcal/mol using the helical. There is a significant variation in the average potentials at the higher $\lambda$ values, with $\lambda = 0.8$ showing a standard deviation of $\pm 35$~kcal/mol and $\lambda = 1.0$ showing a standard deviation of $\pm 57$~kcal/mol [figure~\ref{figure2}~(d)]. Thus, the rough estimate of the current approximation suffers from statistical uncertainty in the averages used.

\subsection{Water distribution about peptide}

\begin{figure}[!h]
\centering
\includegraphics[width=0.51\textwidth]{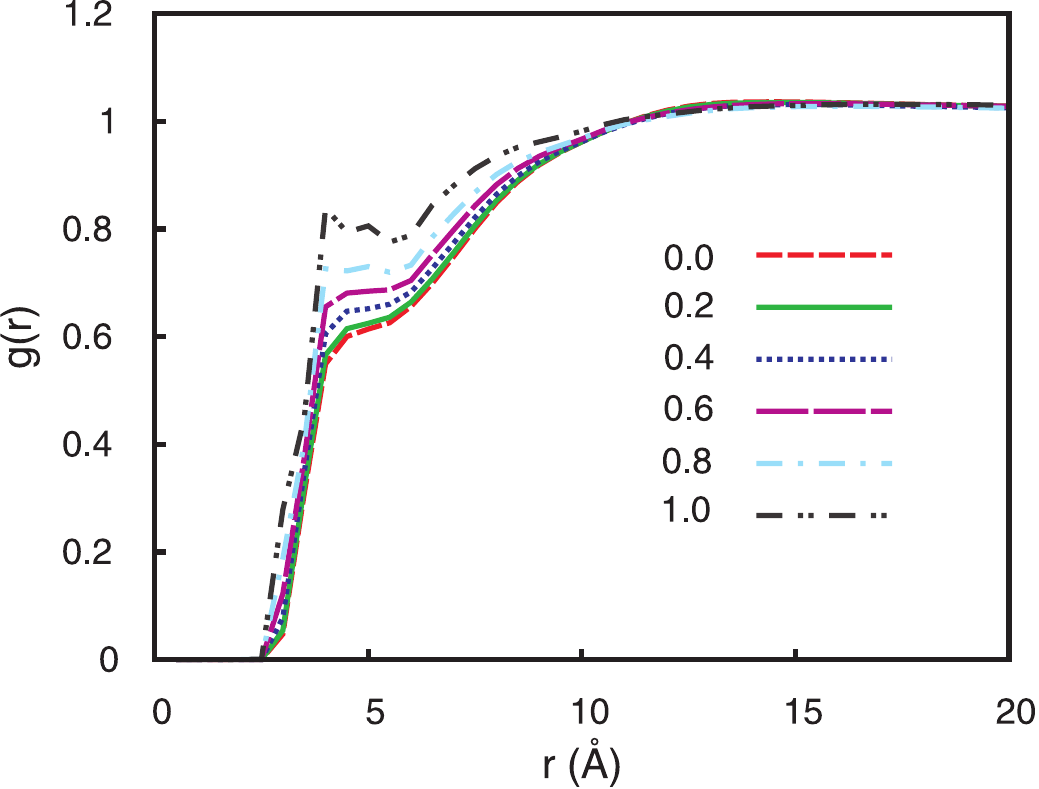}
\caption[Radial distribution of water about peptide with respect to $\lambda$.]{(Color online) Radial distribution of water oxygens about non-hydrogen peptide atoms for the different $\lambda$ windows. The color scheme for each $\lambda$ is described in the key.}
\label{figure3}
\end{figure}

In order to examine the consequences of the competition between the electrostatics and peptide-water vdW potentials as reflected in the solvation in greater detail, we calculated the radial distribution of water oxygen atoms about all peptide heavy atoms for all $\lambda$ windows. The results are shown in figure~\ref{figure3}. The contact peak shifts in all the curves are a reflection of interatomic interactions. The induced structure is reminiscent of the charging process in other polar fluids \cite{Hirata1982, Pettitt1982, Pettitt1983}. At lower $\lambda$ values, these small distance peaks appear as shoulders on the curve, but as charges are increased, they become more pronounced and even split into two peaks at $\lambda = 1.0$  due to the formation of peptide-water hydrogen bonds. Beyond $\sim 6$~\AA, the curve depicts water distribution with respect to the conformationally averaged peptide. At $\lambda = 0.0$, only van der Waals interactions occur between peptide and water molecules (as well as within the peptide). Thus, the location of the first peak reflects the effective vdW  distance, i.e., the formation of the solvent cavity. As charges are turned on, water forms the expected electrostatic interactions with the peptide that allows the aspects of the cavity size to decrease, lowering the distance at which the first peak occurs as $\lambda$ values increase. We also see the effect of the increase in the radius of gyration as the charge increases in the tails of the distributions of the radius of gyration.

\section{Discussion}

The peptide-water electrostatic potential becomes increasingly favorable as charges are turned on. The approximate electrostatic solvation free energy for Gly$_{15}$ is quite favorable, as has been observed for other, shorter oligoglycines \cite{Hu2010b, Staritzbichler2005}. However, in spite of the large favorable solvation free energy, oligoglycines collapse due to the formation of favorable intrapeptide interactions. Here, oligoglycine in each $\lambda$ window collapsed versus a random chain, with the most compact manifold of structures occurring at $\lambda = 0.0$. As $\lambda$ increases, the peptide explores larger conformations as it forms favorable electrostatic interactions with water and has both short-ranged internal repulsions and longer-ranged hydrogen bonding opportunities. Few intramolecular H-bonds result, however, typically less than 3 for Gly$_{15}$.

\begin{figure}[!h]
\centering
\includegraphics[width=0.45\textwidth]{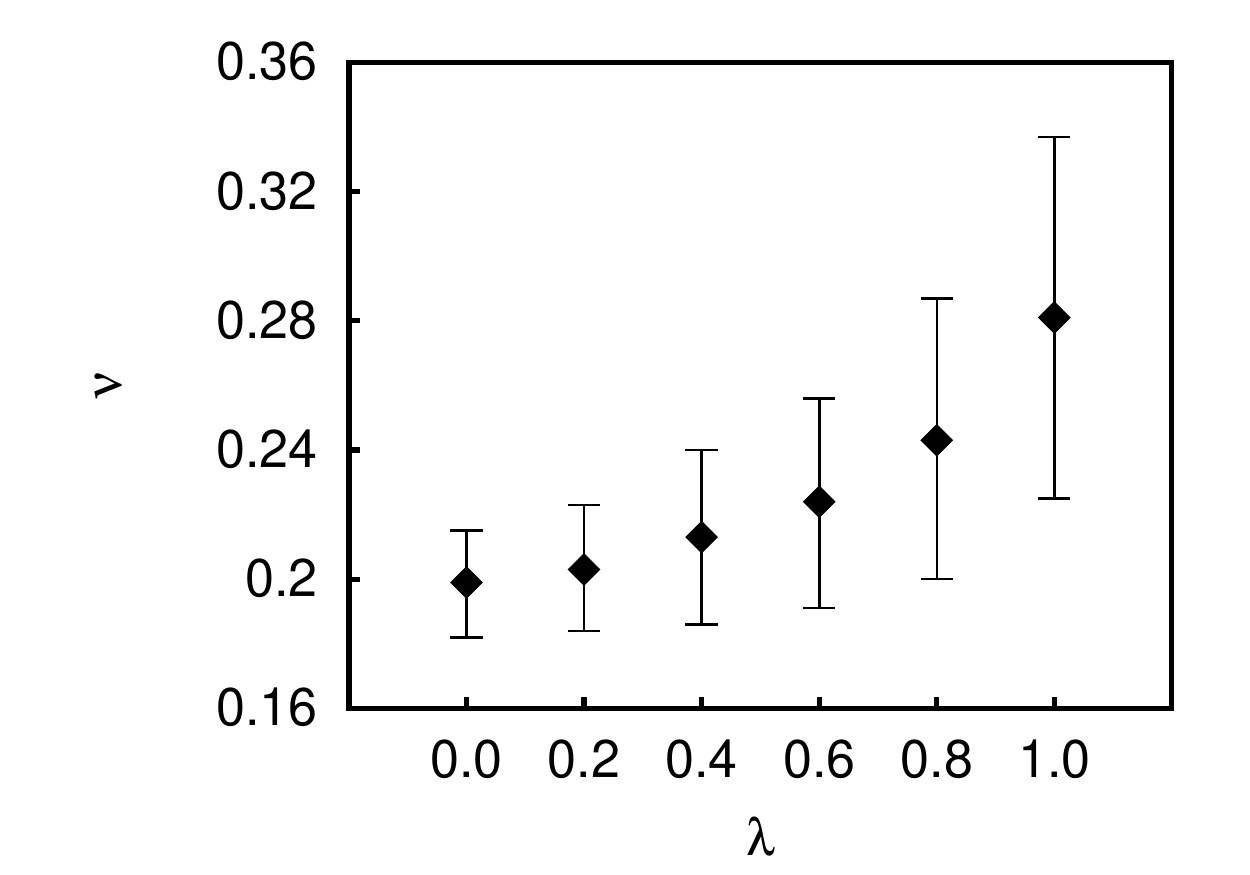}
\caption{The radius of gyration exponent $\nu$ with respect to $\lambda$. The error bars demonstrate the variability of $R_\text{g}$.}
\label{figure4}
\end{figure}

From polymer theory, the degree of collapse can be quantified by the following relationship:
\begin{equation}
R_\text{g} = aN^{\nu},
\end{equation}
where $R_\text{g}$ is the radius of gyration, $a$ is the length of a monomer unit (3~{\AA} in peptides, obtained from the oligoglycine peptide in the fully extended state), $N$ is the number of monomer units (15 in this case) and $\nu$ is a scaling exponent \cite{Flory1953}. For a polymer in a ``good'' solvent $\nu \cong  0.6$, for a polymer in a ``poor'' solvent $\nu \cong  0.3$ and in a  solvent where the polymer is expected to behave like an ideal linear chain, $\nu  = 0.5$. The rare conformations with the largest $R_\text{g}$ were observed when $\lambda = 1.0$. The mean $R_\text{g}$ for $\lambda = 1.0$ is 6.5~{\AA}~--- the $R_\text{g}$ which is expected in a relatively poor solvent (see figure~\ref{figure4}).
Thus, while the peptide forms interactions with water as charges are turned on, and the peptide-water interaction potential becomes increasingly favorable, the peptide does not form kinetically stable extended conformations, but only briefly explores these conformations before collapsing again due to the balance of forces.

%

While peptides collapse due to cavity forces and favorable intrapeptide vdW interactions when the charges are turned off, these interactions become  less dominant as charges are turned on. Some of the change in vdW interactions is due to the inevitable tension between the vdW repulsions and the electrostatic attractions. The decreasing favorability of the total intrapeptide potential with an increasing $\lambda$ is caused in part by a repulsion between the oxygen atoms from adjacent CO-CO dipoles oriented parallel to each other as observed previously \cite{Karandur2016}. When these interactions are discounted, electrostatic intrapeptide interactions become favorable and compete with peptide-water interactions to stabilize the collapsed conformations. Since intrapeptide H-bonds have been observed with  low probability in oligoglycines \cite{Hu2010a, Karandur2014a, Karandur2016}, these interactions are primarily non-hydrogen bonded interactions between amide dipoles, or the so-called CO-CO interactions. The configurationally averaged interactions between the CO-CO pairs are favorable after removing the constrained neighbor oxygen repulsions. Furthermore, it has been observed during simulations of aggregation of pentaglycines that while the intrapeptide electrostatic potential was positive during aggregation, the interpeptide electrostatic potential was negative and decreased continuously during aggregation \cite{Karandur2014a}. The intrapeptide potential in a pentaglycine would come primarily from the neighbouring CO dipoles whereas the interpeptide potential comes from CO dipoles forming favourable interactions.


\section{Conclusions}

In this study, we simulated the Gly$_{15}$ in water while varying the peptide charge. Peptides in all the charge windows collapsed, even with the strongly favorable electrostatic solvation free energy. The extent of collapse depended on the charge, with peptides having lower charge collapsing to a greater extent due to favorable intrapeptide vdW interactions. However, when charges are turned on, the peptides formed favorable intrapeptide interactions, especially when the interactions between oxygen atoms of adjacent CO dipoles were discounted, which overcame the large electrostatic solvation free energy of the system. We have previously shown that while H-bonds are absent in collapsed oligoglycines, the interactions between CO dipoles helped stabilize the collapsed state \cite{Karandur2016}.

\section*{Acknowledgements}

The Robert A. Welch Foundation (H-0037), the National Science Foundation (CHE-1152876) and the National Institutes of Health (GM-037657) are thanked for partial support of this work. This research was performed in part using the Keenland and Stampede systems, part of the National Science Foundation XSEDE resources.


%
%

\ukrainianpart

\title{Внесок електростатичних взаємодій у колапс \\ олігогліцину у воді
}
\author{Д. Карандур\refaddr{label1}, Б.М. Петтітт\refaddr{label1,label2}}
\addresses{
\addr{label1}Структурна і обчислювальна біологія та молекулярна біофізика, Коледж медицини Бейлора, \\ Г'юстон, Техас 77030, США
\addr{label2} Центр Сейлі структурної біології і молекулярної біофізики, Факультет біохімії і молекулярної біології, Медичне відділення Університету Техасу,  Університетський бульвар, 301, Ґалвестон, \\ Техас 77555-0304, США
}

\makeukrtitle

\begin{abstract}
\tolerance=3000%
Розчинність та конформаційна стабільність протеїну є результатом балансу взаємодій як в межах протеїну, так і між протеїном та розчинником. Вільна енергія електростатичної сольватації олігогліцинів, моделей для хребта протеїну, стає більш вигідною із зростанням довжини, до того ж довші пептиди колапсують через формування вигідних внутріпептидних взаємодій між диполями CO, в деяких випадках без водневих зв'язків. Сильно відштовхувальне формування розчинникової порожнини збалансовується ван~дер~вальсівським притяганням та електростатичними внесками. Для того, щоб дослідити конкуренцію між виключенням розчинника та зарядовими взаємодіями, ми моделюємо колапс довгого олігогліцину з 15-ма блоками при скейлінгу зарядів на пептиді від нуля до повного заряду. Ми вивчаємо, який ефект це має на конформаційні властивості пептиду. Ми також описуємо приблизні термодинамічні зміни, що відбуваються під час скейлінгу як через інтрапептидні потенціали, так і через потенціали пептид-вода, і визначаємо вільну енергію електростатичної сольватації для системи.

\keywords вільна енергія гідратації, колапс олігогліцину

\end{abstract}

\end{document}